\documentclass[%
 reprint,
superscriptaddress,
 amsmath,amssymb,
 aps,
]{revtex4-2}

\usepackage{graphicx}
\usepackage{dcolumn}
\usepackage{bm}
\usepackage{amssymb,amstext,amsmath}
\usepackage{xr}
\usepackage{color}

\definecolor{darkred}{rgb}{0,0,0}
\definecolor{darkblue}{rgb}{0,0,0}
\newcommand{\blue}[1]{\textcolor{darkblue}{#1}}
\newcommand{\red}[1]{\textcolor{darkred}{#1}}

\begin{document}

\preprint{APS/123-QED}

\title{Evolutionary game dynamics for higher-order interactions}

\author{Jiachao Guo}
\affiliation{%
Center for Systems and Control, College of Engineering, Peking University, Beijing 100871, China}
\author{Yao Meng}
\affiliation{%
Center for Systems and Control, College of Engineering, Peking University, Beijing 100871, China}
\author{Aming Li}
\thanks{Corresponding author: amingli@pku.edu.cn}
\affiliation{%
Center for Systems and Control, College of Engineering, Peking University, Beijing 100871, China}
\affiliation{
Center for Multi-Agent Research, Institute for Artificial Intelligence, Peking University, Beijing 100871, China}



\date{\today}

\begin{abstract}

  Cooperative behaviors are deeply embedded in structured biological and social systems. Networks are often employed to portray \red{pairwise interactions} among individuals, where network \red{nodes represent individuals and links indicate who interacts with whom}. However, it is increasingly recognized that many \red{empirical} interactions \red{often} involve triple or more individuals instead of \red{the massively oversimplified lower-order} pairwise interactions, highlighting the fundamental gap in understanding the evolution of collective cooperation for higher-order interactions \red{with diverse scales of the number of individuals}. Here, we \red{develop} a theoretical framework \red{of evolutionary game dynamics} for systematically analyzing how cooperation evolves and fixates under higher-order interactions. Specifically, we offer a simple condition under which cooperation is favored under arbitrary combinations of different orders of interactions. Compared to pairwise interactions, our findings \red{suggest} that higher-order interactions enable lower thresholds for the emergence of cooperation. \red{Surprisingly}, we show that higher-order interactions \red{favor the evolution} of cooperation in large-scale \red{systems, which is the opposite for lower-order scenarios}. Our results offer a new avenue for understanding the evolution of \red{collective} cooperation in empirical systems with higher-order interactions.

\end{abstract}

\maketitle

\section{Introduction}

\begin{figure*}
	\centering
	\includegraphics[width=0.95\textwidth]{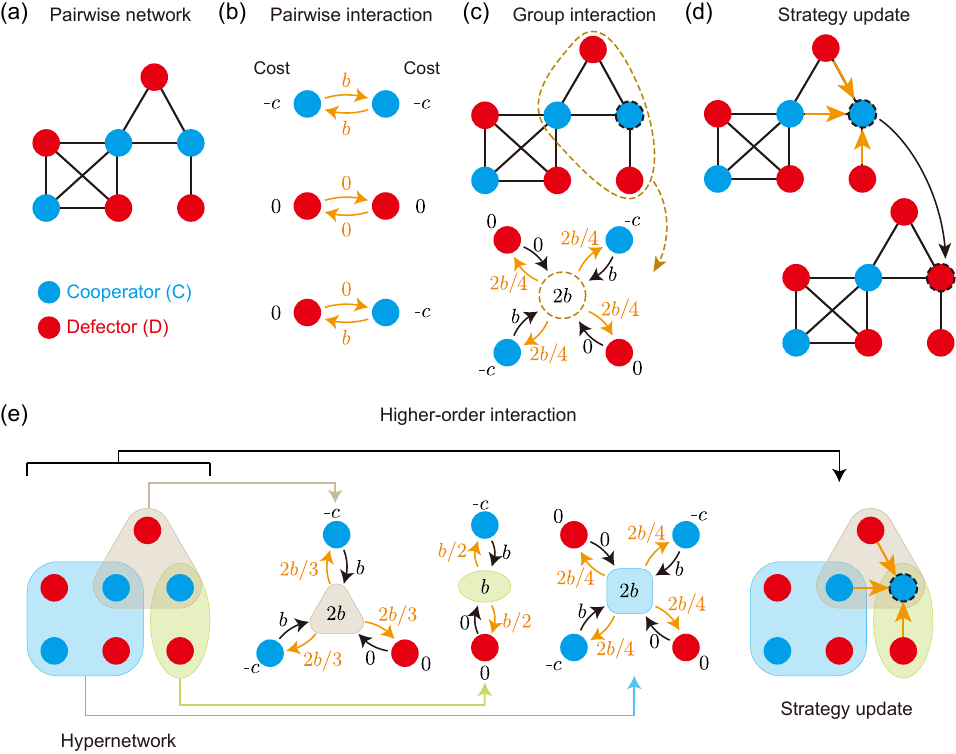}
	\caption{Illustration of the evolutionary process.
		 (a) In the pairwise network, an edge can connect only two individuals. (b) Pairwise interactions are based on such edges. Each individual plays the donation game with their neighbors and accumulates benefits from all games. (c) Group interaction is also based on pairwise networks, where individuals form a group with their neighbors to play a public goods game. (d) The strategy updates for these two interactions are based on pairwise networks where individuals learn the strategy of a particular neighbor. (e) Higher-order interactions involve multiple individuals interacting simultaneously, and hypernetworks can effectively characterize such interactions by allowing edges, known as hyperedges, to connect multiple individuals.
		 In the hypernetwork, each hyperedge corresponds to a public goods game.
		 When performing strategy updates for higher-order interactions, the focal individual learns the strategies of individuals who have shared hyperedges with the focal individual. Indeed, if an individual shares more hyperedges with a particular neighbor, there is a higher probability of learning that neighbor’s strategy.
		}
	\label{fig:figure1}
\end{figure*}

Collective cooperation is one of the central pillars in the evolution of various species, from cellular organisms to human societies. Researchers have long sought to understand why and how individuals often cooperate for mutual benefits \cite{smith_evolution_1982,sigmund2010calculus, nowak2004emergence}. Social dilemma captures the essence of this phenomenon \cite{axelrod1981evolution, nowak1992evolutionary}, in which cooperators pay a cost to provide benefits to others for the highest social welfare, while defectors do not contribute anything for the sake of personal optimization.
Network reciprocity [Fig. \ref{fig:figure1}(a)] is an important mechanism for promoting collective cooperation in social dilemmas \red{\cite{nowak2006five, ohtsuki2007direct, su2023strategy}, where the network} provides a restricted local interaction range to facilitate the formation of clusters of cooperators for mutual benefits to resist invasions of defectors. Indeed, complex \red{networks offer} an efficient way to capture structured populations \cite{barabasi2013network, taylor2007evolution, debarre2014social, hauert2004spatial, perc2013evolutionary, li2020evolution, meng2024dynamics, wang2023imitation}, where nodes represent individuals and edges denote pairwise interactions between each pair of nodes [Fig. \ref{fig:figure1}(b)].
According to the paradigm of pairwise interactions, extensive research offers significant insights into the evolution of \red{collective} cooperation \cite{santos2005scale, ohtsuki_simple_2006, allen_evolutionary_2017,allen_mathematical_2019,mcavoy_social_2020,mcavoy_fixation_2021, tarnita_evolutionary_2009}.
The classical simple rule indicates that, cooperation can emerge if the ratio between the benefit provided by a cooperator and the associated cost paid exceeds the average number of neighbors when all individuals have a similar number of neighbors \cite{ohtsuki_simple_2006}. In more general cases, Allen \textit{et al}. provide the critical ratio, above which cooperation is promoted for arbitrary networks \cite{allen_evolutionary_2017}.

But cooperation in real-world scenarios is not only a dyadic phenomenon where each pair of individuals interact over an edge pairwisely \cite{levine_beyond_2017, battiston_networks_2020, rosenheim1998higher, mickalide2019higher, kugler2012groups}. Specifically, traditional pairwise interactions do not encapsulate the true essence of collective cooperation in the whole population \red{for} group interactions \red{with more individuals}. And they are better described by higher-order interactions involving more than two individuals simultaneously \cite{mayfield2017higher}, which transcend the scope of simple combinations of the lower-order pairwise interactions.
Recently, it \red{is} reported that hypernetworks [Fig. \ref{fig:figure1}(e)] serve as a more effective model for characterizing higher-order interactions \cite{johnson2013hypernetworks}. They extend the concept of traditional pairwise networks by allowing edges, known as hyperedges, to connect more individuals.
Indeed, researchers underscore that the implementation of hypernetworks can significantly promote cooperation, as evidenced by multiple studies \cite{gomez2011evolutionary, alvarez-rodriguez_evolutionary_2021, wang2024evolutionary, fotouhi_evolution_2019}. These studies predominantly concentrate on the evolution of the frequency of cooperators and analytically calculate the critical threshold, above which cooperation \red{may} emerge.

However, a quantitative and systematical understanding of the emergence from a single to full cooperators on arbitrary hypernetworks remains unclear. The underlying mechanism propelling the evolutionary process of cooperation on hypernetworks still eludes intuitive comprehension. Furthermore, how to promote cooperation on hypernetworks is still an open problem.
Here we develop a theoretical framework for studying fixation dynamics on arbitrary hypernetworks and identify simple critical conditions under which collective cooperation is favored. By revealing the process of strategy dissemination on hypernetworks, we find that higher-order interactions can mitigate the negative effects of an increasing number of individual neighbors, thereby promoting the emergence of cooperation.
Our investigation into higher-order fan models reveals the \red{advantage} of higher-order interactions to foster cooperation in large-scale groups, paving the way for practical implementations \red{to} enhance collective cooperation in diverse settings.

\section{Modeling framework}

\begin{figure*}
	\centering
	\includegraphics[width=0.99\textwidth]{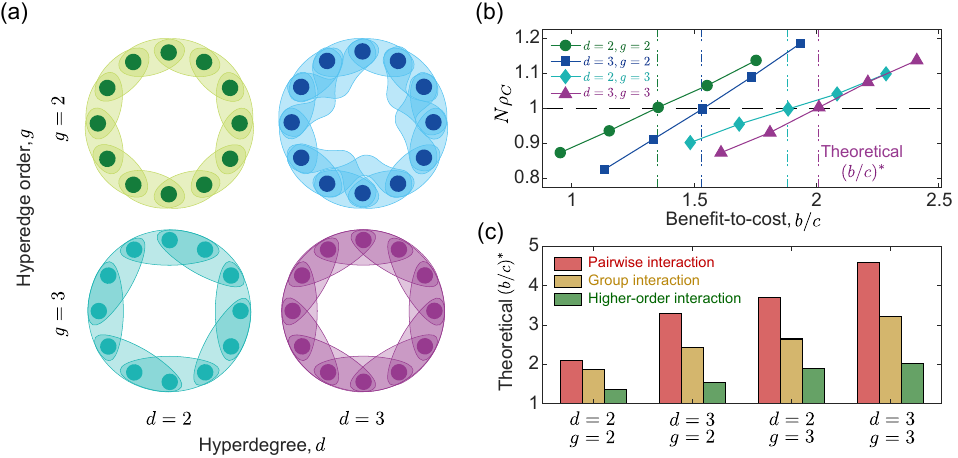}
	\caption{Verification of theoretical predictions. (a) For the verification of the theoretical predictions, we take these four typical hypernetworks as examples. (b) We present the product of the fixation probability of cooperation and the number of individuals ($N\rho_C $) as a function of the benefit-to-cost ratio ($b/c$) under death-birth update across various hypernetworks depicted in (a). The simulation conditions are set as follows: $N=48$, $\delta=0.01$, $c=1$, with the number of numerical simulation iterations set to $5\times 10^6$. Inclined scatter points represent the results of the numerical simulation. Vertical dashed lines indicate theoretical predictions.
	(c) We calculate the critical thresholds of these four hypernetworks for the three interactions described in Fig. \ref{fig:figure1}.
	}
	\label{fig:figure2}
\end{figure*}

We consider \red{the} general evolutionary process in a finite population of $N$ individuals. Individuals may participate in pairwise interactions and interactions within groups \red{at diverse scales} involving more individuals simultaneously, known as higher-order interactions. 
Hypernetworks can effectively characterize higher-order interactions, where each node represents an individual and each hyperedge \red{indicates} a set (group) of individuals interacting simultaneously [Fig. \ref{fig:figure1}(e)]. The node hyperdegree $d$ \red{is defined as} the number of hyperedges an individual (node) is involved into, and the order of hyperedge $g$ \red{is defined as} the number of nodes that form the hyperedge.

\red{Following the convention, here we employ} the public goods game to model the social dilemma involving two or more concurrent individuals \cite{alvarez-rodriguez_evolutionary_2021}.
For game strategy, each individual can be either a cooperator (C) or a defector (D). Cooperators contribute a cost $c$ to a public pool, while defectors do \red{nothing}. The total contributions are then multiplied by an enhancement factor $R=b/c$, and the resultant benefit is distributed equally among all members.
Suppose there are $g_C~(0 \leq g_C \leq g)$ cooperators in a hyperedge which has order $g$, the payoff for a cooperator and a defector can be expressed as $P_C=g_Cb/g-c$ and $P_D=g_Cb/g$, respectively. The total payoff $u_i$ of each individual $i$ is the sum of the payoff from the games in which it participates.

In the evolutionary process, the strategies of the individuals evolve with each iteration of the public goods game.
Widely used strategy evolution rules include \red{death-birth, birth-death, and imitation updating} \cite{allen_evolutionary_2017}. In the main text, we focus on death-birth updating \red{(results for other updating rules are presented in Supplementary Note 1), where an individual is randomly selected to learn} the strategy of one of its neighbors $j$ (individuals that are connected via at least one hyperedge), with probability proportional to the fitness of individual $j$, typically defined as $F_j=1+\delta u_j$. Here $\delta$ is called the intensity of selection \cite{ohtsuki_simple_2006,allen_evolutionary_2017}, which captures how strongly fitness influences the propensity to learn an individual strategy. 
For comparison with existing findings, we focus on the scenario of weak selection ($0<\delta \ll 1$).

To quantitatively describe the evolutionary dynamics of cooperation, we study the emergence of cooperation by comparing the fixation probability of cooperator ($\rho_C$). It is defined as the probability that a single cooperator in a population of $N-1$ defectors generates a lineage of cooperators that does not become extinct but instead takes over the whole population \cite{ohtsuki_simple_2006,allen_evolutionary_2017,allen2019evolutionary,allen_mathematical_2019}. If the fixation probability of the cooperator exceeds $1/N$, then we say selection favors cooperators replacing defectors.

\section{Results}

\subsection{Conditions for the  evolutionary success of cooperation}
To intuitively understand the evolutionary \red{game} dynamics of \red{collective} cooperation \red{with higher-order interactions}, we first take four \red{typical} configurations of hypernetworks as examples [Fig. \ref{fig:figure2}(a)]. For these hypernetworks, they \red{exhibit} different critical conditions for the emergence of cooperation, \red{which is manifested by} different critical benefit-to-cost \red{ratio} $(b/c)^*$. As illustrated in Fig. \ref{fig:figure2}(b), when the benefit-to-cost exceeds the critical value $(b/c)^*$, cooperation is favored, namely, $N\rho_C>1$.
It is worth noting that an increase \red{of} the hyperdegree and hyperedge order leads to an increase in the critical value. This suggests that these factors play a crucial role in determining the conditions for the emergence of cooperation \red{with higher-order interactions}.

Having completed the numerical groundwork, we then transition to theoretical predictions of the critical conditions.
\blue{To describe the strategy updating on hypernetworks, here we propose a \red{new method by mapping hypernetwork to the traditional pairwise replacement network}, based on the number of hyperedges shared between individuals. And \red{we prove} that the strategy updating on the hypernetwork is equivalent to that on the replacement network (Supplementary Note 1).}
\red{We further} shed light on our numerical results by deriving a closed-form expression for the critical benefit-to-cost ratio $(b/c)^*$ as a function of the hypernetwork structure (Supplementary Note 1)
\begin{equation}
	\begin{aligned}
		(b/c)^* = \frac{\sum_{i,j}\pi_i p_{ij}^{(2)}d_j\eta_{ij}}{\sum_{i,j,k}\pi_i p_{ik}^{(2)}\widetilde{w}_{kj}\eta_{ij} - \sum_{i,j}\pi_i \widetilde{w}_{ij}\eta_{ij}}
	\end{aligned}.
	\label{eq:main_1}
\end{equation}
Here, $\pi_i$ captures the reproductive value of the individual $i$, and $d_i$ denotes the hyperdegree of individual $i$. The probability of a 1-step ($n$-step) random walk from $i$ to $j$ is denoted by $p_{ij}^{(1)}$ ($p_{ij}^{(n)}$), and $\eta_{ij}$ denotes the coalescence time—the expected time for two random walks starting from nodes $i$ and $j$ to meet at a common node.
In $\widetilde{w}_{ij}=\sum_\alpha g_{\alpha}^{-1}h_{i\alpha}h_{j\alpha}$, the term $g_\alpha$ denotes the order of the hyperedge $\alpha$, and $h_{i\alpha}$ denotes whether individual $i$ belongs to the hyperedge $\alpha$. If individual $i$ belongs to the hyperedge $\alpha$, then $h_{i\alpha}=1$ ($h_{i\alpha}=0$, otherwise). Our theoretical predictions of Eq. (\ref{eq:main_1}) [vertical dotted lines in Fig. \ref{fig:figure2}(b)] \red{align with} the results of the numerical simulation [inclined scatters in Fig. \ref{fig:figure2}(b)].

In addition, we proceed to a comparative analysis of critical conditions for the emergence of cooperation for higher-order interactions, group interactions \cite{santos_social_2008,szolnoki2009topology, li_evolutionary_2016, wang_spatial_2024}, and pairwise interactions, on these four hypernetworks. \red{Group interactions indicate that the focal individual and its pairwise neighbors form a group to play a public goods game [Fig. \ref{fig:figure1}(c)]}. To enable a uniform comparison, we utilize the hypernetwork of higher-order interactions as our baseline. The network skeletons of pairwise and group interactions are constructed by the process of well-mixing within \red{the hyperedges}.
We discover with surprise that among these three paradigms of interaction, higher-order interaction always has the lowest critical threshold, meaning that it is more conducive to cooperation [Fig. \ref{fig:figure2}(c)].

\subsection{A simple rule for the evolution of cooperation on hypernetworks}

\begin{figure*}
	\centering
	\includegraphics[width=0.99\textwidth]{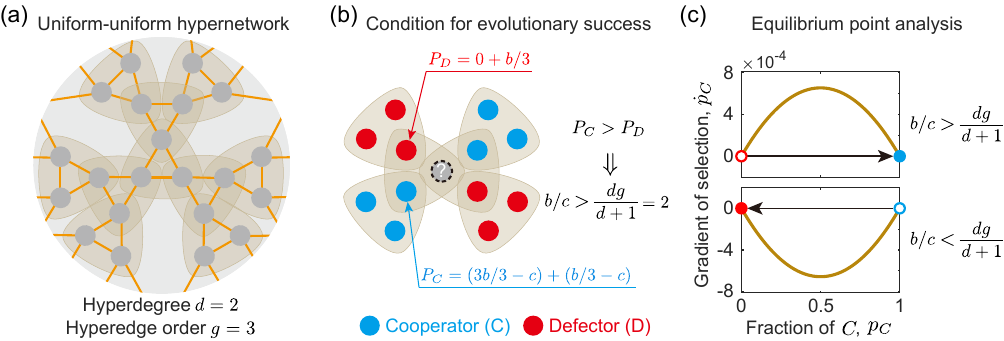}
	\caption{Some intuitive explanations of games on hypernetworks. (a) The pair-approximation is based on this special uniform-uniform hypernetwork that satisfies that all node hyperdegrees are equal and all hyperedge orders are equal. We visualize the hypernetwork and its corresponding replacement network (orange pairwise edges). (b) We consider a focal individual (grey node) selected to update its strategy, it will learn the strategy of one of its neighbors. Pair-approximation calculation shows that for weak selection the cooperator has one more cooperator among its $d(g-1)-1$ other neighbors than the defector. Hence, the focal individual has a higher probability of learning the strategy of a cooperator neighbor, if $b/c>dg/(d+1)$. (c) We present the evolutionary dynamics of an infinite population. Parameters are $\delta=0.01$, $c=1$, $d=3$, $g=4$, $b=3.3$ for the top panel and $b=2.7$ for the bottom panel. The direction of selection dynamics is indicated by the arrow, where the small solid circle represents a stable equilibrium and an empty circle represents an unstable equilibrium. 
		}
	\label{fig:figure3}
\end{figure*}

To \red{further} explore the crucial role of hyperdegree and hyperedge order for the emergence of cooperation \red{on hypernetworks}, we \red{first} consider hypernetworks \red{with} equal node hyperdegree $d$ and hyperedge order $g$ (we call them uniform-uniform hypernetworks [Fig. \ref{fig:figure3}(a)]). 
Using pair-approximation and diffusion approximation (Supplementary Note 2), we find \red{that} cooperation is favored when
\begin{equation}
	\begin{aligned}
		b/c>\frac{dg}{d+1}.
	\end{aligned}
	\label{eq:pairApprox}
\end{equation}
Intuitively, \red{this suggests that} the critical benefit-to-cost $(b/c)^*$ is linear with the hyperedge order $g$ but asymptotically saturating growth with the hyperdegree $d$. This implies that the influence of the hyperedge order is greater than that of the hyperdegree.

We present the intuitive justification for the rule in Fig. \ref{fig:figure3}(b). If a focal individual is selected to update its strategy [empty node in Fig. \ref{fig:figure3}(b)], it will learn the strategy of one of its neighbors according to their payoffs. 
The payoff of the cooperator neighbor of the focal individual consists of two components: the payoff when the cooperator is on the same hyperedge as the focal individual, and the payoff on the other $d-1$ hyperedges. The payoff of the defector neighbor is similar. 
For ease of comprehension, we denote the conditional probability to find a cooperator next to a cooperator as $q_{C|C}$, and to find a cooperator next to a defector as $q_{C|D}$. 
Thus, the payoff of the cooperator neighbor and the defector neighbor of the focal individual can be denoted \red{separately} as 
\begin{equation}
	\begin{aligned}
		P_C&=\left(1+(g-2)q_{C|C}\right)b/g-c \\
       &+(d-1)\left\{\big[\left(1+(g-1)q_{C|C}\right)b\big]/g-c\right\}, \\
		P_D&=(g-2)q_{C|D}b/g+(d-1)(g-1)q_{C|D}b/g.
	\end{aligned}
	\nonumber
\end{equation}
It's worth noting that the payoff that comes from the focal individual (empty node) is excluded, because it contributes equally to the cooperator and the defector.
Hence, the cooperator is favored compared to the defector to disperse its strategy, if $P_C>P_D$. 
Pair-approximation shows that $q_{C|C}-q_{C|D}=1/(d(g-1)-1)$ for weak selection \red{(Supplementary Note 2)}. This implies that the cooperator has on average one more cooperator neighbor than the defector.
Therefore, we obtain $P_C-P_D=b(d+1)-c dg$, which leads to the $b/c>dg/(d+1)$ rule.

Furthermore, we explore the evolution of cooperation in infinite populations [Fig. \ref{fig:figure3}(c)]. We know that in a linear public goods game, the defector's payoff is always greater than that of the cooperator. Therefore, there are only two fixed points, the full cooperator and full defector, and no internal fixed points \cite{gokhale2010evolutionary}.
In terms of the cooperator frequency $p_C$ (Supplementary Note 2), when $b/c>dg/(d+1)$, the full cooperator $p_C=1$ is a stable state, and the full defector $p_C=0$ is unstable. Conversely, $p_C=1$ is unstable and $p_C=0$ is stable. 

To explain why \red{higher-order} interactions have an advantage over the other two interactions [Fig. \ref{fig:figure2}(c)], here, we compare the differences among the results obtained from approximation theory for pairwise interactions, group interactions, and higher-order interactions.
Traditional pairwise interactions decouple complex interactions into linear combinations of two-individual interactions, making the number of neighbors (topological degree) a crucial factor that linearly influences the evolution of cooperation, known as $b/c>k$, where $k$ denotes the number of neighbors \red{\cite{ohtsuki_simple_2006}}. 
In group interactions, an individual participates in $k+1$ public goods games, including that centered on itself and those centered on its $k$ neighbors. 
The latter establishes an association between the individual and its second-order neighbors. 
This leads to the number of neighbors no longer linearly affecting the emergence of cooperation \red{\cite{li_cooperation_2014,li_evolutionary_2016}}, namely, $b/c>(k+1)^2/(k+3)$.

Different from pairwise and group interactions, higher-order interactions based on hypernetworks, by allowing hyperedges to connect more individuals, \red{may} essentially capture this complex nonlinear relationship. Considering the number of neighbors $k=d(g-1)$ in Eq. (\ref{eq:pairApprox}), we discover that the critical threshold converges to hyperedge order $g$ over the number of neighbors $k$. This presents a non-linear relationship between the conditions for the emergence of cooperation and the number of neighbors. It reveals that higher-order interactions can effectively mitigate the negative effects of the increasing number of neighbors, resulting in easier conditions for the emergence of cooperation.

\subsection{From pairwise to higher-order interactions: A unifying perspective}

\begin{figure*}
	\centering
	\includegraphics[width=0.99\textwidth]{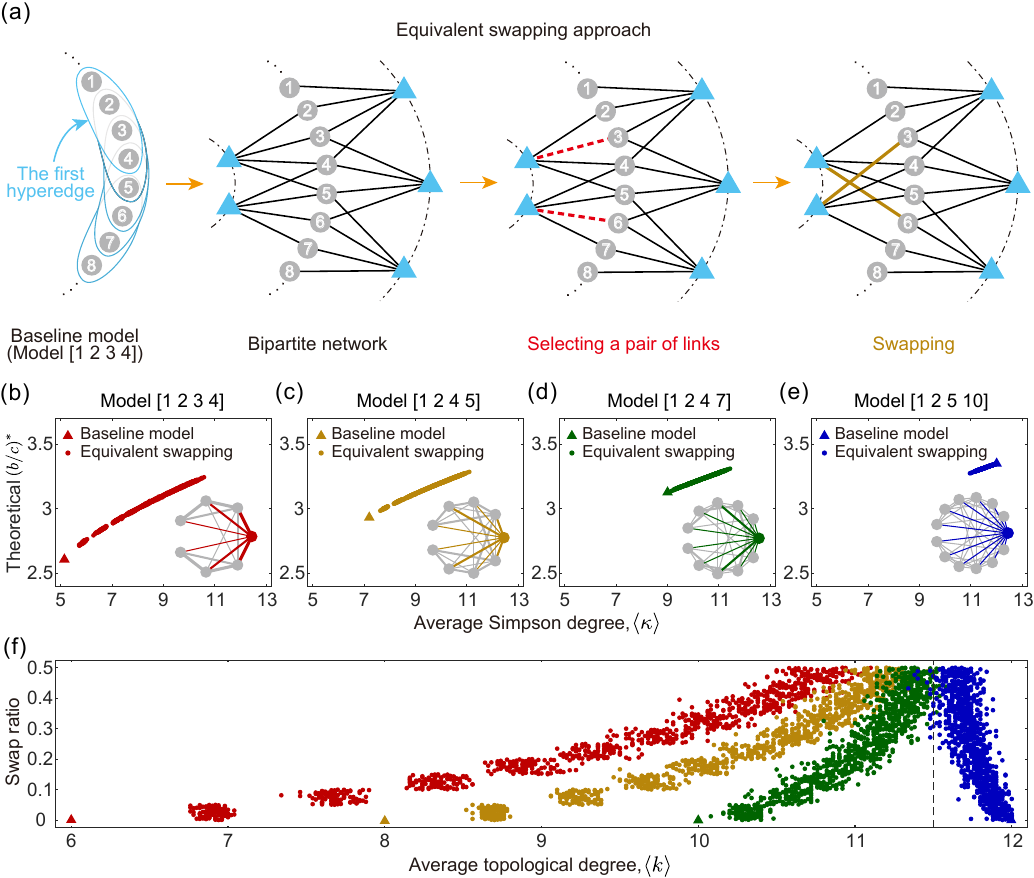}
	\caption{Effect of the Simpson degree on the critical threshold. 
		We number 100 ring-arranged individuals in (a).
		Based on sequential selection approach, the first hyperedge of model ‘[1 2 3 4]’ consists of four individuals numbered 1, 2, 3, 4, the second hyperedge consists of four individuals numbered 2, 3, 4, 5, and continues in this manner. The last hyperedge consists of four individuals numbered 100, 1, 2, and 3. The other three baseline models are similar. 
		Meanwhile, we demonstrate the equivalent swapping approach. The hypernetwork is first represented as a bipartite network. Two links are randomly selected, disconnected, and then exchanged. The proportion of swapped links to total links is defined as the swapped rate, which ranges $ 5\%, 10\%,\cdots, 50\% $.
		In (b)-(e), we show the distributions of the critical values for the average Simpson degree. The four different colors indicate the four baseline models and their corresponding equivalent swapping hypernetworks. The small network in each panel indicates the local replacement network of the baseline model, where the colored nodes indicate the individual we are focusing on. The colored solid lines indicate the edges between the focused node and its neighbors, and the grey solid lines indicate the edges between the neighbors of the focused individual. 
		(f) We present the change in the average topological degree with the equivalent swapping ratio for these four models.
		The simulation conditions are: $ N=100, d=4,g=4$.
		}
	\label{fig:figure4}
	
\end{figure*}

The results of the pair-approximation allow us to explore the advantages of higher-order interactions in terms of the number of neighbors, which in fact provides a potential correlation between the three paradigms of interaction. Considering the limited application scenarios of pairwise approximation, we \red{next} explore more general scenarios. 
Based on our results on arbitrary hypernetworks in Eq. (\ref{eq:main_1}), we consider the case with $d_i=\left<d \right>$, for $i=1,2,\cdots,N$ and $g_{\alpha}=\left<g\right>$, for $\alpha=1,2,\cdots,m$. \red{Here $m$ denotes the number of hyperedges.} This condition suggests that the average hyperdegree $\left<d \right>$ is considered in place of the hyperdegree of each node, and the average order $\left<g\right>$ is utilized instead of the order of each hyperedge.
\red{Accordingly} we derive \red{the extended expression} for Eq. (\ref{eq:main_1}):
\begin{equation}
	\begin{aligned}
		(b/c)^*&=\frac{\left<g\right>\eta^{(2)}}
		{(\left<g\right>-1)(\eta^{(3)}-\eta^{(1)})+\eta^{(2)}},
		\end{aligned}
	\label{eq:main_2}
\end{equation}
where $\eta^{(n)}=\sum_{i,j=1}^N\pi_i p_{ij}^{(n)}\eta_{ij}$ (Supplementary Note 3).
It is well known that $\eta^{(n)}$ is a parameter solely associated with the replacement network, which is obtained from the hypernetwork by mapping. \red{For the donation game on the replacement network}, we obtain the critical benefit-to-cost $\mathcal{B}^* = \eta^{(2)}/(\eta^{(3)}-\eta^{(1)})$ for pairwise interactions \cite{allen_evolutionary_2017}. Considering $\mathcal{B}^*$ in Eq. (\ref{eq:main_2}), we have
\begin{equation}
	\begin{aligned}
		(b/c)^*&
		=\frac{\left<g\right>\mathcal{B}^*}
		{\left<g\right>-1+\mathcal{B}^*}.
		\end{aligned}
	\label{eq:DG}
\end{equation}
It portrays the correlation between the public goods game on the hypernetwork and the donation game on its replacement network. In short, it \red{presents} the correlation between higher-order interactions and pairwise interactions. 

\red{It is natural to} ask how higher-order interactions are related to group interactions.
Here, we first analyze the correlation between group interactions and pairwise interactions on the replacement network of the hypernetwork. \red{We consider} that the replacement network is regular.
Therefore, we obtain the correlation between group interactions and pairwise interactions on arbitrary regular networks
\begin{equation}
	\begin{aligned}
		(b/c)_{\text{Group}}^*&
		=\frac{(k+1)^2\mathcal{B}^*}
		{\mathcal{B}^*+2k+k^2(1+\mathcal{P})},
	\end{aligned}
	\label{eq:group_DG}
\end{equation}
where, $\mathcal{P}=N\mathcal{C}(k-1)/k^3$, and $\mathcal{C}$ is the clustering coefficient of the replacement network \cite{wang2024evolutionary}. If the neighbors of all nodes on the replacement network are not further connected, then the clustering coefficient $\mathcal{C}=0$, such that $\mathcal{P}=0$. \red{On this point, we show that} Eq. (\ref{eq:group_DG}) agrees with the previous research \cite{li_cooperation_2014}. 

By leveraging the foundational correlations established between pairwise and group interactions, \red{we further} explore the intricate correlation between higher-order interactions and group interactions. In this way, we establish strong correlations between the three paradigms of interaction from the perspective of conditions for the emergence of cooperation. Indeed, \red{Eqs. (\ref{eq:DG}) and (\ref{eq:group_DG})} also provide a framework for conversion between different interaction paradigms. This allows us to analyze the emergence of cooperation not only from a single paradigm of interaction but from multiple paradigms of interaction more conveniently and intuitively.

\subsection{General conditions for emergence of cooperation on uniform-uniform hypernetworks}

\red{We know} that even if two hypernetworks have the same statistical distribution of hyperdegree and hyperedge order, different configurations (affiliation of nodes to hyperedges) can significantly impact the emergence of cooperation. This leads us to wonder whether there is a general condition, given hyperdegree distributions and hyperedge order distributions, \red{for promoting cooperation} on all hypernetworks. \red{Next, we consider} the uniform-uniform hypernetworks with symmetry, meaning that the set $\left\{w_{ij}\right\}$ of outgoing weights is the same in the replacement network \cite{allen_evolutionary_2017}. \red{In this case, 
Eq. (\ref{eq:main_2}) is} further simplified as (Supplementary Note 3)
\begin{equation}
		(b/c)^*=\frac{N -2}
		{\frac{g-1}{g}\sum_{i\in G}\frac{1}{\kappa_i}+\frac{N}{g}-2},
		\label{eq:main_3}
\end{equation}
where, $\kappa_i = \big(\sum_{j\in G}p_{ij}^2\big)^{-1}$ is the Simpson degree \cite{allen2019evolutionary}, which indicates the inverse of the sum of squares of one-step transition probabilities from the focal node to its neighbors. 

To analyze the influence of the configuration, we define a `family of uniform-uniform hypernetworks' as a set of uniform-uniform hypernetworks\red{, where they} have different configurations. As examples, we construct four representative uniform-uniform hypernetworks of 100 individuals satisfying $d=4, g=4$, using a sequential selection approach [\red{left panel} in Fig \ref{fig:figure4}(a)]. We call the four hypernetworks as `baseline hypernetworks model', and name them models ‘[1 2 3 4]’, ‘[1 2 4 5]’, ‘[1 2 4 6]’, and ‘[1 2 5 10]’ based on the node numbers of the first hyperedge. 
It is worth noting that other baseline models exist, such as model ‘[1 2 5 6]’, but their configuration is essentially one of these four (Supplementary Note 4).
\red{Furthermore, we offer an equivalent swapping approach for node-hyperedge links based on bipartite networks [Fig. \ref{fig:figure4}(a)] to change configurations of hypernetworks. Specifically, this approach changes the affiliation between nodes and hyperedges, but keeps the distributions of node hyperdegrees and hyperedge orders unchanged, i.e., the new hypernetworks still belong to the `family of uniform-uniform hypernetworks'. We apply the equivalent swapping approach to four baseline hypernetworks to create new hypernetworks.
The swapping ratios studied are $ 5\%, 10\%,\cdots, 50\% $. For each swapping ratio, we calculate the critical threshold $(b/c)^*$  for 100 hypernetworks and collate the results.}

Figures \ref{fig:figure4}(b)-(e) show a correlation between $(b/c)^*$ and $\left<\kappa \right>$.  The threshold $(b/c)^*$ increases with the average of the Simpson degree $\left<\kappa \right>$. Thus, the maximum value of the Simpson degree corresponds to the maximum value of the critical threshold, which coincides with Eq. (\ref{eq:main_3}). According to the arithmetic mean inequality, the maximum value of Simpson degree is equal to the maximum number of topological neighbors (on the replacement network), $k=d(g-1)$. Therefore, we obtain
\begin{equation}
	(b/c)^*_{\mathrm{max}}
	=\frac{(N-2)gd}
	{N(d+1)-2gd}.
	\label{eq:max}
\end{equation}
Indeed, it corresponds to the result of the baseline model ‘[1 2 5 10]’ [blue solid triangle in Fig. \ref{fig:figure4}(e)]. 
In this case, the equivalent swapping approach \red{reduces} the average topological degree [Fig \ref{fig:figure4}(f)], which leads to a lower Simpson degree and critical thresholds. While the equivalent swapping approach \red{boosts} the number of neighbors in the other three models. This is because their replacement networks tend to be an equal-weight replacement network with an average topological degree $\left<k\right>\approx11.5$ [\red{Fig. \ref{fig:figure4}(f)}], as the swapping ratio increases.

In summary, we obtain the maximum value $(b/c)^*_{\mathrm{max}}$ of the critical threshold for uniform-uniform hypernetworks. \red{Thus, when the benefit-to-cost ratio of the altruistic act exceeds the maximum value of the critical threshold, namely, $b/c>(b/c)^*_{\mathrm{max}}$, all hypernetworks in this `family of uniform-uniform hypernetworks' is favorable for cooperation}, which provides new meaningful criteria for evaluating the impact of hypernetworks on the evolution of cooperation.
In larger populations, we obtain the simple rule $(b/c)^*_{\mathrm{max}}=dg/(d+1)$. On the one hand, this implies that pair-approximation generally results in the maximum value on critical thresholds in larger populations. The reason is that on the assumption of approximation, the number of neighbors (topological degree) of the individual is maximal, \red{namely, $k=d(g-1)$}. On the other hand, \red{to make all hypernetworks} in the `family of uniform-uniform hypernetworks' conducive to cooperation, a quick and effective way is satisfying the simple rule for the evolution of cooperation on hypernetworks.

\subsection{Boosting cooperation for large groups with higher-order links}

\begin{figure*}
	\centering
	\includegraphics[width=0.99\textwidth]{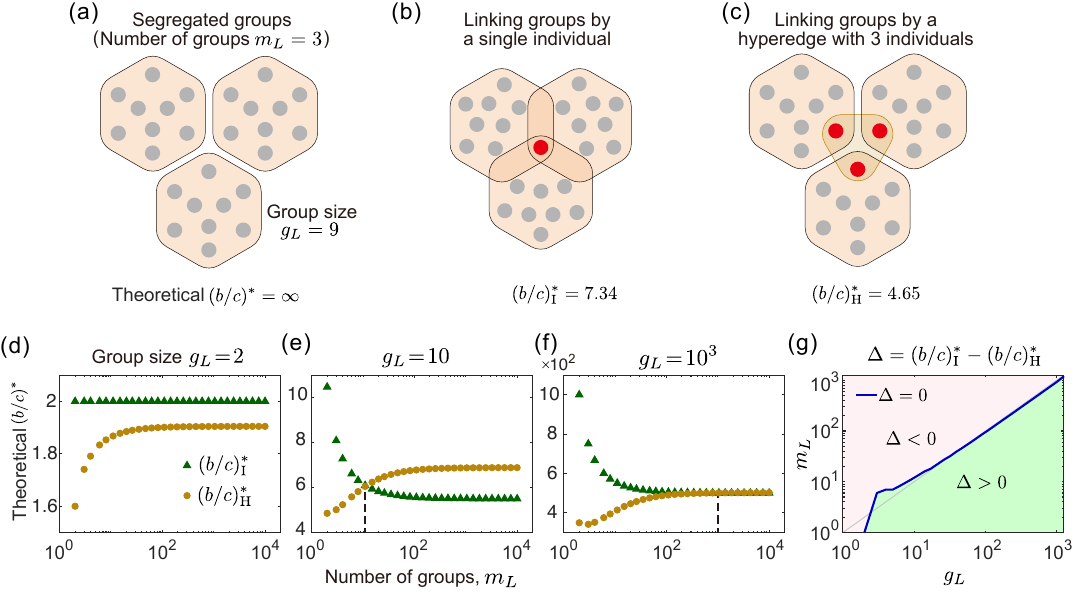}
	\caption{
		The effect of \red{linking} group methods on the evolution of cooperation. (a) We denote three segregated groups of nine individuals represented by the grey nodes. In (b) and (c), we show different ways of \red{linking} groups and the corresponding critical benefit-to-cost $(b/c)^*$ that favors cooperation, respectively. 
    In (b), the three groups are linked by a single individual, denoted by a red node. And in (c), the three groups are linked by an additional hyperedge, indicated by a dark gold solid triangle, consisting of individuals from each group.
    The panels (d)-(f) indicate the trend of the threshold for cases shown in (b) and (c) over different numbers of groups and group sizes, and the vertical dashed lines in (e) and (f) indicate when $(b/c)^*_{\mathrm I}=(b/c)^*_{\mathrm H}$. (g) We plot the magnitude of the threshold, and the blue line indicates $(b/c)^*_{\mathrm I}=(b/c)^*_{\mathrm H}$. The grey solid line indicates that the group size is equal to the number of groups.
	}
	\label{fig:figure5}
	
\end{figure*}

Previous investigations report that unstructured populations do not facilitate cooperation \red{\cite{hofbauer1998evolutionary}}. However, when populations are \red{linked} through nodes or edges, they can significantly enhance cooperative behavior \cite{allen_evolutionary_2017,fotouhi2018conjoining}. This naturally motivates us to explore whether \red{linked} populations \red{by additional links} on a hypernetwork favor cooperation. Furthermore, considering the feature that hyperedges contain multiple individuals, we are interested in determining whether \red{addtional links} based on higher-order interactions are more conducive to cooperation compared to \red{the links} involving a single individual.
Here, we consider a higher-order fan structure, in which the leaves are \red{linked} via a hyperedge [Fig. \ref{fig:figure5}(c)]. We define the method of \red{linking segregated groups} [Fig. \ref{fig:figure5}(a)] by a hyperedge as \red{higher-oder links}. For comparison, we also define a fan structure \red{linked} by a single individual [Fig. \ref{fig:figure5}(b)].

We use $g_L$ to denote the number of groups and \red{$m_L$ the number of members in each group}, i.e., the group size.
In the scenario of segregated groups [Fig. \ref{fig:figure5}(a)], since the network is perfectly symmetric, its Simpson degree $\kappa$ is equal to its topological degree. Additionally, the membership within the group is well-mixed, satisfying $g=N, k=N-1$. When these conditions are satisfied in Eq. (\ref{eq:main_3}), we find that the denominator is equal to 0, hence the threshold $(b/c)^*\to \infty$, indicating that cooperation is never favored.
We calculate the critical threshold for the emergence of cooperation for hypernetworks in Figs. \ref{fig:figure5}(b) and \ref{fig:figure5}(c) using Eq. (\ref{eq:main_1}) (Supplementary Note 5). We find that groups \red{linked} by a hyperedge have smaller critical thresholds. This implies that the higher-order \red{link} is more favorable to cooperation, compared to a single individual \red{linking} multiple groups. However, this finding is not consistent with previous research that larger group sizes always inhibit cooperation.

To explore this counterintuitive result further, we define the advantage of higher-order \red{links} over single-individual \red{links}, $\Delta =(b/c)^*_{\mathrm I}-(b/c)^*_{\mathrm H}$, for varying numbers of groups and group sizes. We select the three most representative scenarios with $g_L=2,10,10^3$, corresponding to group sizes of 2, 10, and 1000, respectively [Figs. \ref{fig:figure5}(d)-(f)]. 
When the group size is 2, the hypernetwork \red{linked} through a single individual is a star network, which has a critical threshold that is constant and higher than that of the higher-order \red{link}. This means that the higher-order \red{link} is always more favorable for cooperation. When the group size increases slowly, i.e., $g_L=10$, the critical benefit-to-cost of the two \red{linking} approaches intersect as the number of groups $m_L$ increases. The number of groups at this intersection is approximately equal to the group size, implying that when the number of groups is smaller than the group size, the higher-order \red{link} is more conducive to cooperation. Conversely, when the number of groups exceeds the group size, the single-individual \red{link} becomes more favorable. As the group size continues to increase, i.e., $g_L=10^3$, the critical benefit-to-cost of both methods converges to an equal constant related to $g_L$, implying the two ways promote cooperation equivalently.

We summarize the conditions under which the higher-order \red{link} is more conducive to cooperation: The group size exceeds or equals the number of groups [the green area in Fig. \ref{fig:figure5}(h)]. 
Indeed, when the group size is large enough, even if the group size is smaller than the number of groups, the critical values of the two ways of \red{linking} are nearly equal [Fig. \ref{fig:figure5}(f)]. We categorize this region as neutral. 
The area where higher-order \red{links} have the advantage of promoting cooperation [green area in Fig. \ref{fig:figure5}(f)] is larger than the area where the single-individual \red{link} promotes cooperation [red area in Fig. \ref{fig:figure5}(f)].
Therefore, higher-order interactions can offer the potential for the emergence of cooperation in large-scale groups.

\section{Discussion}

We have proposed a theoretical framework to systematically analyze the evolution of cooperation within the context of higher-order interactions, revealing the advantages of higher-order interactions in facilitating the emergence of cooperation. 
This framework offers a novel perspective for understanding collective behavior in complex systems. 
Simultaneously, we have delved into the intricate correlations between different paradigms of interaction. This allows us to analyze the emergence of cooperation from multiple perspectives more conveniently and intuitively.

Furthermore, we present a novel and effective criterion for assessing the impact of hypernetworks on the evolution of cooperation. This general criterion can simplify the analysis of how various hyperdegree and hyperedge order distributions influence cooperative evolution.
In our analysis of the cooperative evolution of large-scale populations, we discover that higher-order \red{links} can lower the threshold for promoting the emergence of cooperation, which is unusual, thus making large-scale group cooperation possible.

\red{Our results show} the rich dynamics of cooperative evolution under higher-order interactions. This framework of higher-order interactions holds promising potential for further exploration across various fields. For example, in epidemic modeling \cite{pastor2015epidemic, nowzari2016analysis, pastor2001epidemic, keeling2005networks}, higher-order interactions may reveal the complex transmission pathways among individuals and the mechanisms of herd immunity formation. In information diffusion \cite{moreno2004dynamics}, higher-order interactions may help us better understand the propagation patterns of information in social networks and the influence of key nodes.

A natural extension of our findings is \red{to explore} the scenario with multiple strategies \cite{gokhale2010evolutionary, van_veelen_multi-player_2012, wangchaoqian2024evolutionary}. Existing studies indicate that the implementation of multiple strategies often results in richer dynamic characteristics, such as cyclic dominance captured in the classic rock-paper-scissors game \cite{szolnoki2014cyclic}. \red{The coupling} of multi-strategy with multi-player is necessary when considering more realistic models. Our explorations in the field of two-strategy multi-player interactions provide a theoretical basis for this extension. For example, we can further explore the emergence of cooperation, by incorporating additional strategies such as punishing defectors \cite{fehr2002altruistic, boyd2003evolution} or rewarding cooperators \cite{sigmund2001reward}.

\red{Another} promising direction for future research lies in exploring the evolutionary dynamics on temporal hypernetworks \cite{li2017fundamental, li2020evolution}. \red{In some cases} hypernetworks are not static\red{, and} they evolve over time. Temporal hypernetworks capture not only the interactions among a group of individuals but also the timing of these interactions. Existing studies indicate that temporal pairwise networks often facilitate the emergence of cooperation. Consequently, the coupling of temporal hypernetworks with fixation dynamics may lead to more intricate and exotic evolutionary dynamics. By treating a temporal hypernetwork as a series of static snapshots, our theoretical framework can be applied to each static snapshot to calculate the fixation probability, thereby providing theoretical support for the evolution of cooperation across the entire timeline.

\bibliography{reference}

\end{document}